# Quantum circuit-like learning:
# A fast and scalable classical machine-learning algorithm with similar performance to quantum circuit learning

**Naoko Koide-Majima**[a,b]†, **Kei Majima**[c,d,e]†

[a] Center for Information and Neural Networks (CiNet),
National Institute of Information and Communications Technology, Osaka 565-0871, Japan
[b] Graduate School of Frontier Biosciences, Osaka University, Osaka, 565-0871, Japan
[c] Graduate School of Informatics, Kyoto University, Kyoto 606-8501, Japan
[d] Institute for Quantum Life Science, National Institutes for Quantum Science and Technology, Chiba 263-8555, Japan
[e] JST PRESTO, Saitama 332-0012, Japan
† These authors contributed equally to this work.

The application of near-term quantum devices to machine learning (ML) has attracted much attention. Recently, Mitarai et al. (2018) proposed a framework to use a quantum circuit for ML tasks, called quantum circuit learning (QCL). Due to the use of a quantum circuit, QCL employs an exponentially high-dimensional Hilbert space as its feature space. However, its efficiency compared to classical algorithms remains unexplored. Here, we present a classical ML algorithm that uses the same Hilbert space. In numerical simulations, our algorithm demonstrates similar performance to QCL for several ML tasks, providing a new perspective for the computational and memory efficiency of quantum ML algorithms.

## I. INTRODUCTION

Because quantum computers with tens or hundreds of qubits are becoming available, their application to machine learning has attracted much attention. To use noisy intermediate-scale quantum (NISQ) devices [1] efficiently, many quantum-classical hybrid algorithms have been proposed. For example, the variational quantum eigensolver (VQE) has recently been used to find the ground state of a given Hamiltonian [2–5] and the quantum approximate optimization algorithm (QAOA) enables us to obtain approximate solutions to combinatory optimization problems [6–8]. More recently, several studies have proposed algorithms that use NISQ devices for machine-learning tasks [9–29], some of which have been experimentally tested using actual quantum devices [30–36].

In one such attempt, Mitarai et al. [26] proposed a framework to train a parameterized quantum circuit for supervised classification and regression tasks; this is called quantum circuit learning (QCL). In QCL, input data (*i.e.*, feature vectors) are nonlinearly mapped into a $2^Q$-dimensional Hilbert space, where $Q$ is the number of qubits, and then a parameterized unitary transformation is applied to the mapped data (see Section II). The parameters of the unitary transformation are tuned to minimize a given cost function (*i.e.*, a prediction error). According to the original QCL study [26], QCL is considered to have four unique properties as a machine-learning model:

1) Nonlinear mapping into the high-dimensional Hilbert space provides QCL with the ability to learn nonlinear relationships.

2) The unitary transformation maintains the norm of the weight vector (*i.e.*, coefficient vector) at one, which mitigates the risk of overfitting.

3) The weight vector is randomly parameterized.

4) The parameters of the weight vector can be optimized using a gradient method with analytical gradients.

While these four properties lead to advantages in practical machine-learning tasks, it remains uncertain whether they are exclusive features of quantum machine learning against classical machine learning.

In this paper, we present a classical machine-learning algorithm with the above four properties whose computational time and required memory size are linear with respect to the hyperparameter corresponding to the number of qubits, which we refer to as quantum circuit-like learning (QCLL). To implement QCLL, we introduce the technique known as "count sketch" [37–40], which is a randomized algorithm for linear algebra computation. Given a high-dimensional vector space, the count sketch technique provides a projection from the given space to a lower dimensional space that approximately preserves the inner product in the original space (see Section II). Therefore, this enables us to use the same $2^Q$-dimensional Hilbert space as the feature space with a low computational cost. To demonstrate the similarities between QCL and QCLL as machine-learning algorithms, we perform numerical simulations in which these two algorithms are applied to several machine-leaning tasks. Our numerical results demonstrate that the behavior and performance of QCLL are very similar to those of QCL.

Specifically, the contributions of our study are as follows:

- We proposed a classical machine-learning algorithm that uses a $2^Q$-dimensional Hilbert space as its feature

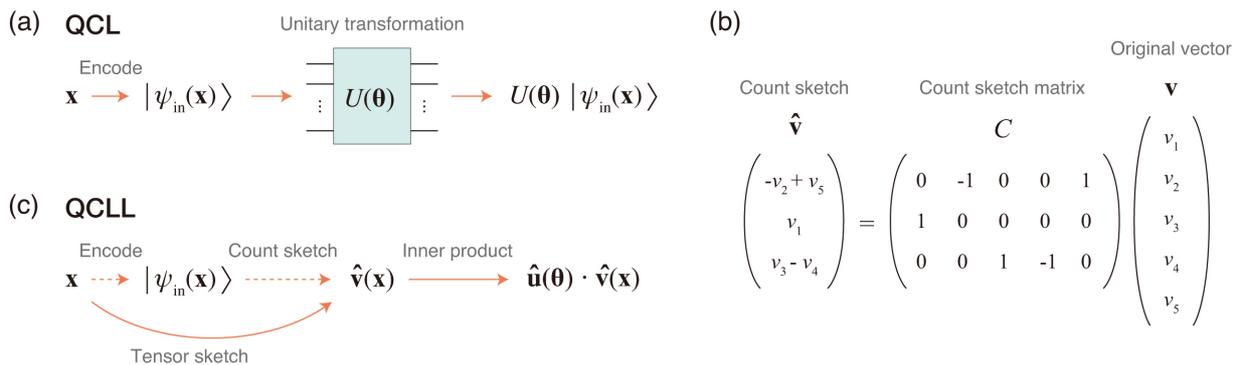

**FIG. 1.** Quantum circuit learning (QCL) and quantum circuit-like learning (QCLL). (a) Flowchart of QCL. In QCL, an input vector is encoded (mapped) into a quantum state and then a parameterized unitary transformation is applied to it. (b) Count sketch technique. An example of a 3 × 5 count sketch matrix is shown. This provides a projection from a five-dimensional space to a three-dimensional space. (c) Flowchart of QCLL. In QCLL, the count sketch of the quantum state vector that encodes an input vector is computed by the tensor sketch algorithm. The tensor sketch algorithm enables us to perform this computation without accessing the quantum state vector, which is typically high dimensional. Then, the inner product between the resultant count sketch and a parameterized unit vector is computed as the output.

space while keeping the computational time and required memory size linear with respect to $Q$.

- To efficiently compute the inner product in the $2^Q$-dimensional Hilbert space, several statistical techniques developed in the field of machine learning are introduced. We combined them with the gradient-based optimization procedure developed in the previous QCL study [26].

- We demonstrated that our proposed algorithm achieves prediction performance similar to QCL on the machine-learning tasks used in the original QCL study [26]. This similarity was also confirmed through practical benchmark tasks in the field of machine learning.

While we numerically demonstrate that QCLL shows similar prediction performance to QCL, it should be noted that our results are consistent with the expectations from previous quantum machine learning studies. Several theoretical studies have shown that the difference in learning efficiency (*i.e.*, the number of training samples required to achieve a given prediction performance) between quantum and classical machine-learning algorithms is at most polynomial [41–44]. Therefore, theoretically, for any given quantum machine-learning algorithm, there exists a classical machine-learning algorithm with similar prediction performance. Further discussion on this point is provided in Section IV.

The remainder of this paper is organized as follows: The algorithms used in this study are described in section II. Section III then presents the experimental results. Discussion and concluding remarks are given in Section IV.

## II. ALGORITHM

This section is organized as follows. First, we introduce QCL based on the work of Mitarai et al. [26]. In the following subsection, we explain the count sketch technique, which is a randomized algorithm for linear algebra computation and is used as a subroutine in QCLL. Finally, we explain QCLL. Note that a Python implementation of QCLL is available at our GitHub repository [https://github.com/nkmjm/].

### A. Quantum circuit learning

Here, we introduce the QCL framework proposed by Mitarai et al. [26]. In supervised learning, an algorithm is given a training dataset $\mathcal{D} = \{(\mathbf{x}_n, y_n)\}_{n=1}^N$ of $N$ pairs of $D$-dimensional real input vectors $\mathbf{x}_n \in \mathbb{R}^D$ and target outputs $y_n$. The outputs $y_n$ take continuous/discrete values in the case of regression/classification. The algorithm is required to predict the target output for a new input vector $\mathbf{x}_{\text{new}} \in \mathbb{R}^D$. In QCL, the prediction model is constructed using a quantum circuit with multiple learnable parameters as follows (see also Figure 1(a)).

1. Encode input vectors $\mathbf{x}_n$ ($n = 1, \cdots, N$) into quantum states $|\psi_{\text{in}}(\mathbf{x}_n)\rangle = U_{\text{enc}}(\mathbf{x}_n)|0\rangle^{\otimes Q}$ by applying unitary gates $U_{\text{enc}}(\mathbf{x}_n)$ to the initialized state $|0\rangle^{\otimes Q}$.

2. Apply a $\boldsymbol{\theta}$-parameterized unitary $U(\boldsymbol{\theta})$ to the states $|\psi_{\text{in}}(\mathbf{x}_n)\rangle$ and generate the output states $|\psi_{\text{out}}(\mathbf{x}_n, \boldsymbol{\theta})\rangle = U(\boldsymbol{\theta})|\psi_{\text{in}}(\mathbf{x}_n)\rangle$.

3. Measure the expectation values of a predefined observable $B$ for the output states $|\psi_{\text{out}}(\mathbf{x}_n, \boldsymbol{\theta})\rangle$. Using a predefined function $F(\cdot)$, a scaling parameter $a$, and an intercept parameter $b$, the predictions from this machine-learning model are defined as

$$\hat{y}_n = F(a\langle\psi_{\text{out}}(\mathbf{x}_n,\boldsymbol{\theta})|B|\psi_{\text{out}}(\mathbf{x}_n,\boldsymbol{\theta})\rangle + b). \qquad (1)$$

4. Minimize the cost function
$$C(a,b,\boldsymbol{\theta}) = \sum_{n=1}^{N} E(y_n, \hat{y}_n) \qquad (2)$$
with respect to the parameters $(a,b,\boldsymbol{\theta})$, where $E(\cdot,\cdot): \mathbb{R} \times \mathbb{R} \to \mathbb{R}$ is a function to measure the prediction error.

5. Make a prediction for a new input vector $\mathbf{x}_{\text{new}}$ by computing
$$F(a\langle\psi_{\text{out}}(\mathbf{x}_{\text{new}},\boldsymbol{\theta})|B|\psi_{\text{out}}(\mathbf{x}_{\text{new}},\boldsymbol{\theta})\rangle + b). \qquad (3)$$

In step 1 above, the input vectors are nonlinearly mapped into a $2^Q$-dimensional Hilbert space, where $Q$ is the number of qubits used for encoding. In Mitarai et al. [26], the unitary gates $U_{\text{enc}}(\mathbf{x}_n)$ were constructed using rotations of individual qubits around the y-axis. Following their procedure, in this study, we treat a case in which a given input vector $\mathbf{x} = (x_1, \cdots, x_D)^T \in \mathbb{R}^D$ is encoded in
$$|\psi_{\text{in}}(\mathbf{x})\rangle = \otimes_{d=1}^{D}\left\{\otimes_{q_d=1}^{Q_d}\left(x_d, \sqrt{1-x_d^2}\right)^T\right\}, \qquad (4)$$
where $Q_d$ is the number of qubits used to encode the $d$-th element of the vector $\mathbf{x}$. Note that $Q = \sum_{d=1}^{D} Q_d$. The elements of the vector $|\psi_{\text{in}}(\mathbf{x})\rangle$ include high-order polynomials such as $x_d^{Q_d}$; this introduces nonlinearity into the prediction model.

In step 2, a parameterized unitary transformation $U(\boldsymbol{\theta})$ is applied to the states prepared in step 1. In Mitarai et al. [26], this unitary transformation was constructed as a chain of rotations of individual qubits and random unitary transformations. Following their procedure here, we assume a case where $U(\boldsymbol{\theta})$ takes the form
$$U(\boldsymbol{\theta}) = R_M(\boldsymbol{\theta})U_M R_{M-1}(\boldsymbol{\theta})U_{M-1}\cdots R_1(\boldsymbol{\theta})U_1, \qquad (5)$$
where $M$ is the depth of the quantum circuit, $R_1(\boldsymbol{\theta}), \cdots, R_M(\boldsymbol{\theta})$ are rotations of individual qubits whose angles are specified by elements of $\boldsymbol{\theta}$, and $U_1, \cdots, U_M$ are random unitary transformations. As a result, each element of the output vector (*i.e.*, the output state) $|\psi_{\text{out}}(\mathbf{x},\boldsymbol{\theta})\rangle$ has the form of the inner product between $|\psi_{\text{in}}(\mathbf{x})\rangle$ and a unit vector randomly parameterized by $\boldsymbol{\theta}$. In other words, when we denote $|\psi_{\text{in}}(\mathbf{x})\rangle$ and $|\psi_{\text{out}}(\mathbf{x},\boldsymbol{\theta})\rangle$ by
$$\mathbf{v}_{\text{in}}(\mathbf{x}) = \left(v_{\text{in},1}(\mathbf{x}), \cdots, v_{\text{in},2^Q}(\mathbf{x})\right)^T$$
and
$$\mathbf{v}_{\text{out}}(\mathbf{x},\boldsymbol{\theta}) = \left(v_{\text{out},1}(\mathbf{x},\boldsymbol{\theta}), \cdots, v_{\text{out},2^Q}(\mathbf{x},\boldsymbol{\theta})\right)^T,$$
each element of $|\psi_{\text{out}}(\mathbf{x},\boldsymbol{\theta})\rangle$ has the form

$$v_{\text{out},i}(\mathbf{x},\boldsymbol{\theta}) = \mathbf{u}_i(\boldsymbol{\theta}) \cdot \mathbf{v}_{\text{in}}(\mathbf{x}) \quad (i = 1, \cdots, 2^Q), \qquad (6)$$

where $\mathbf{u}_i(\boldsymbol{\theta})$ is a unit vector randomly parameterized by $\boldsymbol{\theta}$.

The predefined function $F(\cdot)$ and the loss function $E(\cdot,\cdot)$ used in steps 3 and 4 are manually defined. In a regression task, $F(x) = x$ and $E(y,\hat{y}) = (y-\hat{y})^2$ are often used. In a classification task, the softmax and cross-entropy functions are often used. To minimize the cost function $C(a,b,\boldsymbol{\theta})$, gradient-based methods can be used [26], which was also adopted in our numerical simulation.

### B. Count sketch technique

To approximately compute the inner product in the $2^Q$-dimensional Hilbert space in an efficient manner on classical computers, we use a technique called "count sketch." In this section, we assume that we want to approximate the inner product of $K$-dimensional vectors.

**Definition 1**
*Given two independent hash functions $h: \{1, \cdots, K\} \to \{1, \cdots, K'\}$ and $s: \{1, \cdots, K\} \to \{+1, -1\}$, the $K' \times K$ matrix whose $(k', k)$-entry is*

$$\begin{cases} s(k) & (h(k) = k') \\ 0 & (\text{otherwise}) \end{cases} \qquad (7)$$

*is called a count sketch matrix.*

Note that, because $h$ and $s$ are hash functions with randomness, the count sketch matrix specified by $h$ and $s$ is a random matrix. According to convention, a sample (*i.e.*, an observation) drawn from a count sketch matrix is also called a count sketch matrix in this study; however, the interpretation is clear from the context. An example of a count sketch matrix is shown in Figure 1(b).

**Definition 2**
*Given a $K' \times K$ count sketch matrix $C$, the count sketch of a $K$-dimensional column vector $\mathbf{v}$ is defined as the $K'$-dimensional vector $C\mathbf{v}$.*

Count sketches are random vectors. According to convention, the random vector $C\mathbf{v}$ and a sample (*i.e.*, an observation) from $C\mathbf{v}$ are both called the count sketch of $\mathbf{v}$.

Count sketch matrices have the property shown in Theorem 3; this allows us to use count sketches as compact representations of high-dimensional vectors.

**Theorem 3**
*We assume that $C$ is a $K' \times K$ count sketch matrix. Given two $K$-dimensional column vectors $\mathbf{v}_1$ and $\mathbf{v}_2$,*

$$\mathbb{E}_C[C\mathbf{v}_1 \cdot C\mathbf{v}_2] = \mathbf{v}_1 \cdot \mathbf{v}_2, \qquad (8)$$

$$\text{Var}_C[C\mathbf{v}_1 \cdot C\mathbf{v}_2] \leq \frac{1}{K'}\{(\mathbf{v}_1 \cdot \mathbf{v}_2)^2 + \|\mathbf{v}_1\|^2\|\mathbf{v}_2\|^2\}, \quad (9)$$

where $\mathbb{E}_x[f(x)]$ and $\text{Var}_x[f(x)]$ denote the expectation and variance, respectively, of a function $f(x)$ with respect to a random variable $x$.
Proof: See [37,39].

Using the above property, we can approximate $\mathbf{v}_1 \cdot \mathbf{v}_2$ by computing $C\mathbf{v}_1 \cdot C\mathbf{v}_2$. Once we obtain $C\mathbf{v}_1$ and $C\mathbf{v}_2$, this computation can be performed in $O(K')$ time because $C\mathbf{v}_1$ and $C\mathbf{v}_2$ are $K'$-dimensional vectors. This computational time is independent of the original dimensionality $K$ of the space. Note that, because vectors we primarily treat in later sections are unit vectors, the variance (*i.e.*, the approximation error) is always smaller than $2/K'$ in such a case. Based on the numerical results in a previous study [37], $K'$ was set to 100 in our numerical experiments.

In QCL, we treat vectors that take the form $\mathbf{v} = \otimes_{q=1}^Q \mathbf{v}_q$ (see Eq. (4)). In the following, we describe an algorithm to compute the count sketch of a vector with this form in a computationally efficient manner.

**Theorem 4**
*We assume that a $D_{\text{prod}}$-dimensional vector $\mathbf{v}$ can be represented by*

$$\mathbf{v} = \otimes_{q=1}^Q \mathbf{v}_q, \quad (10)$$

*where $\mathbf{v}_q$ ($q = 1, \cdots, Q$) are $D_q$-dimensional vectors, and we assume that $C$ and $C_q$ are $K' \times D_{\text{prod}}$ and $K' \times D_q$ count sketch matrices, respectively. The random vector $C\mathbf{v}$ and the convolution of the random vectors $C_1\mathbf{v}_1, \cdots, C_Q\mathbf{v}_Q$ follow the same distribution.*
Proof: See [38].

Because the convolution of vectors can be computed using a fast Fourier transform (FFT), $C\mathbf{v}$ in the above can be computed as follows.

1. Compute $C_q\mathbf{v}_q$ ($q = 1, \cdots, Q$), where $\mathbf{v}_q$ are given $D_q$-dimensional vectors and $C_q$ are $K' \times D_q$ count sketch matrices.

2. Compute $\text{FFT}(C_q\mathbf{v}_q)$ using FFT.

3. Compute $\text{FFT}(C_1\mathbf{v}_1) \odot \cdots \odot \text{FFT}(C_Q\mathbf{v}_Q)$ where $\odot$ denotes the element-wise product.

4. Compute $\text{FFT}^{-1}\left(\text{FFT}(C_1\mathbf{v}_1) \odot \cdots \odot \text{FFT}(C_Q\mathbf{v}_Q)\right)$. Return this as the output of the algorithm.

According to convention, we call this algorithm "tensor sketch" in this paper. Note that the computation of $C_q\mathbf{v}_q$ ($q = 1, \cdots, Q$) in step 1 of the tensor sketch algorithm takes $O(K'(D_1 + D_2 + \cdots + D_Q))$ computational time. This is smaller than $O(K'D_1D_2\cdots D_Q)$, which is the time for direct computation, and the computational time for steps 2–4 does not depend on the dimensionality of $\mathbf{v}_q$ (*i.e.*, $D_q$).

Among variants of the FFT algorithm, we used the Cooley–Tukey algorithm for our numerical experiments, which exactly computes the discrete Fourier transform of a given sequence without approximation.

### C. Quantum circuit-like learning

Using the count sketch technique and the tensor sketch algorithm, we present a machine-learning algorithm with similar properties to QCL. As in QCL, we assume that the algorithm is given a training dataset $\mathcal{D} = \{(\mathbf{x}_n, y_n)\}_{n=1}^N$. The hyperparameter corresponding to the number of qubits used for encoding the $d$-th element of the input vectors is denoted by $Q_d$. The QCLL algorithm is as follows (see also Figure 1(c)).

1. Draw $Q = \sum_{d=1}^D Q_d$ samples from a $K' \times 2$ count sketch matrix, and compute the count sketches of $\mathbf{v}_{\text{in}}(\mathbf{x}_n) = \otimes_{d=1}^D \left\{\otimes_{q=1}^{Q_d} \left(x_{nd}, \sqrt{1 - x_{nd}^2}\right)^{\text{T}}\right\}$ using the tensor sketch algorithm. Denote the resultant count sketches by $\hat{\mathbf{v}}_{\text{in}}(\mathbf{x}_n)$.

2. Draw $P$ samples from a $K' \times 2$ count sketch matrix, and compute the count sketch of $\mathbf{u}(\boldsymbol{\theta}) = \otimes_{p=1}^P \left(\cos\theta_p, \sin\theta_p\right)^{\text{T}}$ using the tensor sketch algorithm. Repeat this $I$ times, and denote the resultant count sketches by $\hat{\mathbf{u}}_i(\boldsymbol{\theta})$ ($i = 1, \cdots, I$). Then, compute $\hat{v}_{\text{out},i}(\mathbf{x}_n, \boldsymbol{\theta}) = \hat{\mathbf{u}}_i(\boldsymbol{\theta}) \cdot \hat{\mathbf{v}}_{\text{in}}(\mathbf{x}_n)$, and denote them collectively by $\hat{\mathbf{v}}_{\text{out}}(\mathbf{x}_n, \boldsymbol{\theta}) = \left(\hat{v}_{\text{out},1}(\mathbf{x}_n, \boldsymbol{\theta}), \cdots, \hat{v}_{\text{out},I}(\mathbf{x}_n, \boldsymbol{\theta})\right)$.

3. Using a predefined Hermitian matrix $B$, compute $\hat{\mathbf{v}}_{\text{out}}(\mathbf{x}_n, \boldsymbol{\theta})^\dagger B \hat{\mathbf{v}}_{\text{out}}(\mathbf{x}_n, \boldsymbol{\theta})$. Using a predefined function $F(\cdot)$, a scaling parameter $a$, and an intercept parameter $b$, the predictions from this machine-learning model are defined as

$$\hat{y}_n = F(a\hat{\mathbf{v}}_{\text{out}}(\mathbf{x}_n, \boldsymbol{\theta})^\dagger B \hat{\mathbf{v}}_{\text{out}}(\mathbf{x}_n, \boldsymbol{\theta}) + b). \quad (11)$$

4. Minimize the cost function

$$C(a, b, \boldsymbol{\theta}) = \sum_{n=1}^N E(y_n, \hat{y}_n) \quad (12)$$

with respect to the parameters $(a, b, \boldsymbol{\theta})$, where $E(\cdot, \cdot): \mathbb{R} \times \mathbb{R} \to \mathbb{R}$ is a function to measure the prediction error.

5.  Make a prediction for a new input vector $\mathbf{x}_{\text{new}}$ by computing

$$F(a\hat{\mathbf{v}}_{\text{out}}(\mathbf{x}_{\text{new}}, \boldsymbol{\theta})^\dagger B \hat{\mathbf{v}}_{\text{out}}(\mathbf{x}_{\text{new}}, \boldsymbol{\theta}) + b). \quad (13)$$

In step 1 of the above algorithm, $2^Q$-dimensional vectors $\otimes_{d=1}^{D} \left\{ \otimes_{q=1}^{Q_d} \left( x_{nd}, \sqrt{1-x_{nd}^2} \right)^T \right\}$ are encoded as the $K'$-dimensional vectors $\hat{\mathbf{v}}_{\text{in}}(\mathbf{x}_n)$. According to Theorem 3, the values of the norm of $\hat{\mathbf{v}}_{\text{in}}(\mathbf{x}_n)$ are 1.0 on average and, for an arbitrary pair of input vectors $\mathbf{x}_{n_1}$ and $\mathbf{x}_{n_2}$, $\hat{\mathbf{v}}_{\text{in}}(\mathbf{x}_{n_1}) \cdot \hat{\mathbf{v}}_{\text{in}}(\mathbf{x}_{n_2})$ is equal to $\mathbf{v}_{\text{in}}(\mathbf{x}_{n_1}) \cdot \mathbf{v}_{\text{in}}(\mathbf{x}_{n_2})$ on average. This indicates that the pairwise similarities between input vectors in the $2^Q$-dimensional Hilbert space are approximately preserved in the $K'$-dimensional space constructed by this step.

In step 2, each element of $\hat{\mathbf{v}}_{\text{out}}(\mathbf{x}_n, \boldsymbol{\theta})$ takes the same form as in Eq. (6) of QCL. The $i$-th element of $\hat{\mathbf{v}}_{\text{out}}(\mathbf{x}_n, \boldsymbol{\theta})$ can be written as

$$\hat{v}_{\text{out},i}(\mathbf{x}_n, \boldsymbol{\theta}) = \hat{\mathbf{u}}_i(\boldsymbol{\theta}) \cdot \hat{\mathbf{v}}_{\text{in}}(\mathbf{x}_n) \ (i=1,\cdots,I). \quad (14)$$

According to Theorem 3, on average, $\hat{\mathbf{u}}_i(\boldsymbol{\theta})$ is a unit vector randomly parameterized by $\boldsymbol{\theta}$, which has the same property as $\mathbf{u}_i(\boldsymbol{\theta})$ in Eq. (6) of QCL. It should be noted that, because randomly sampled different count sketch matrices are used to produce $\hat{\mathbf{u}}_i(\boldsymbol{\theta})$ and $\hat{\mathbf{v}}_{\text{in}}(\mathbf{x}_n)$, $\hat{\mathbf{v}}_{\text{out}}(\mathbf{x}_n, \boldsymbol{\theta})$ is generally not a separable state vector. To introduce more flexible random parameterization into QCLL, one can optionally insert a randomly sampled unitary matrix in Eq. (14) as follows:

$$\hat{v}_{\text{out},i}(\mathbf{x}_n, \boldsymbol{\theta}) = \hat{\mathbf{u}}_i(\boldsymbol{\theta})^\dagger R \hat{\mathbf{v}}_{\text{in}}(\mathbf{x}_n) \ (i=1,\cdots,I), \quad (15)$$

where $R$ is a random unitary matrix. For simplicity, we adopted Eq. (14) in our main experiments because almost the same results were observed in the comparison between QCLL using Eq. (14) and that using Eq. (15) (see APPENDIX for details). Further discussion on the comparison between Eqs. (14) and (15) is given in Section IV.

Note that steps 3–5 of QCLL are completely the same as those of QCL if we replace $\hat{\mathbf{v}}_{\text{out}}(\cdot,\cdot)$ with $|\psi_{\text{out}}(\cdot,\cdot)\rangle$. Using the tensor sketch algorithm, we can perform QCLL in $O(Q)$ computational time.

Similar to QCL, we can use gradient-based methods to minimize the cost function in step 4 of QCLL. To combine QCL with gradient-based methods, inspired by Li et al [45], Mitarai et al. [26] proposed a method to calculate $\partial \hat{y}_n / \partial \theta_p$. Similar to their approach, in QCLL, we can calculate $\partial \hat{v}_{\text{out},i}(\mathbf{x}_n, \boldsymbol{\theta})/\partial \theta_p$ by evaluating $\hat{v}_{\text{out},i}(\mathbf{x}_n, \boldsymbol{\theta}+\Delta\boldsymbol{\theta})$, where $\Delta\boldsymbol{\theta}$ is a vector whose $p$-th element is $\pi/2$ and whose other elements are zero. By combining this and the chain rule, we can calculate the derivative of the cost function in step 4 of QCLL.

## III. RESULTS

In this section, we demonstrate that the behavior and prediction performance of QCLL are similar to those of QCL using numerical simulations. In the first subsection, we compared QCL and QCLL on the machine-learning tasks used in the original QCL study [26]. In the second subsection, the algorithms were compared using representative benchmark datasets from the field of machine learning. In the final subsection, we conducted a complementary analysis to evaluate the ability of QCLL to approximate QCL.

### A. Behavior for simulation data

Following the numerical simulations in Mitarai et al. [26], we treat the same regression and classification tasks using QCL and QCLL. First, we performed a regression analysis where the functions $f(x) = x^2$, $e^x$, $\sin x$, and $|x|$ were learned (estimated) from a given training dataset $\mathcal{D} = \{(\mathbf{x}_n, y_n)\}_{n=1}^{100}$. Following the procedure in Mitarai et al. [26], the input vectors $\mathbf{x}_n$ are one-dimensional, and were randomly sampled from the range $[-1, 1]$. The target outputs $y_n$ were set to $f(\mathbf{x}_n)$. The number of qubits ($Q$) was set to six, and the depth of the quantum circuit ($M$) was set to six. A matrix whose first five diagonal entries are one and whose other entries are zero was used as the predefined Hermitian matrix $B$. As the parameterized unitary matrix $U(\boldsymbol{\theta})$ of QCL, we used the form of Eq. (5) with

$$R_m(\boldsymbol{\theta}) = \otimes_{q=1}^{6} \begin{pmatrix} \cos(\theta_{6(m-1)+q}) & -\sin(\theta_{6(m-1)+q}) \\ \sin(\theta_{6(m-1)+q}) & \cos(\theta_{6(m-1)+q}) \end{pmatrix}$$
$$(m = 1,\cdots,M)$$

and random unitary matrices $U_m$. Using the procedure applied in a previous study [46], we sampled $U_m$ from the unitary group U($2^6$) uniformly by orthonormalizing complex vectors whose components were sampled from the standard normal distribution. The number of learnable parameters was exactly the same for QCL and QCLL. In other words, $\boldsymbol{\theta}$ in both QCL and QCLL is a 36-dimensional vector in this numerical simulation. $F(x) = x$ and $E(y, \hat{y}) = (y - \hat{y})^2$ were adopted as the predefined function $F(\cdot)$ and the loss function $E(\cdot,\cdot)$, respectively. To minimize the cost function, we used a gradient-based method, SLSQP [47]. The number of iterations was set to 100. To avoid local minima of the cost function, for the same training data, we repeated the SLSQP algorithm 10 times with different initializations and the parameters $(a, b, \boldsymbol{\theta})$ showing the lowest cost function value was used for the prediction. The results are shown in Figure 2. The outputs of QCL and QCLL were fitted to the target functions with similar degrees.

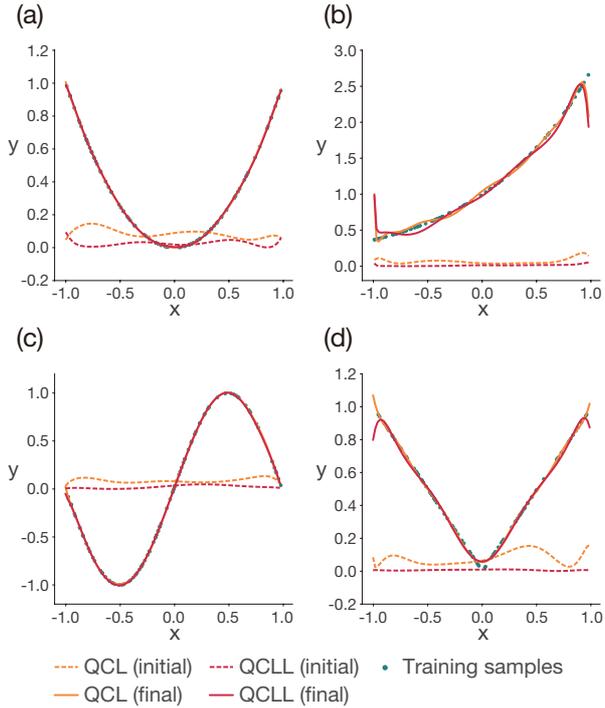
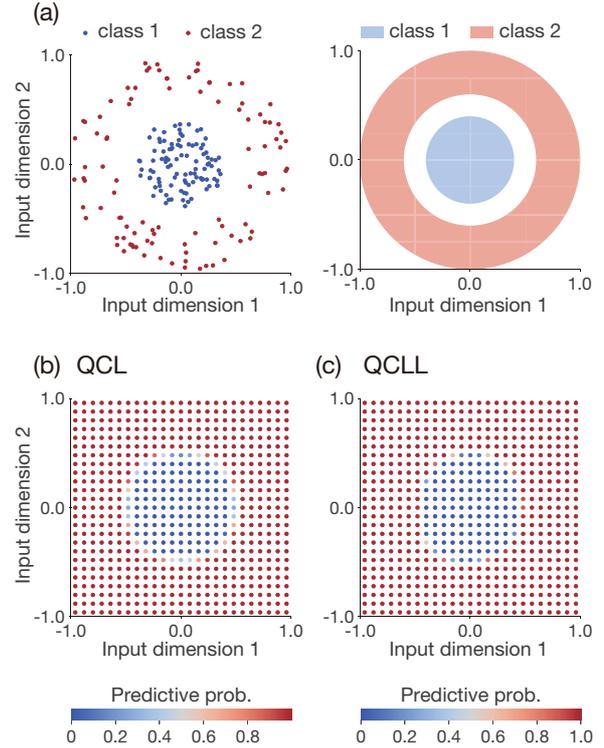

**FIG. 2.** Demonstration of the performances of QCL and QCLL on regression tasks. As regression tasks, four functions (a) y = x$^2$, (b) y = e$^x$, (c) y = sin x, and (d) y = |x| were learned (estimated). Outputs from QCL (orange) and QCLL (red) prior to training (dashed line) and after training (solid line) are shown together with the training samples (green dots).

**FIG. 3.** Demonstration of the performances of QCL and QCLL on a classification task. (a) Training data in the input feature space. The blue and red dots in the left panel indicate the training samples for classes 1 and 2, respectively. These training data were uniformly sampled from the colored areas in the right panel. (b) Predictive probability from QCL after training. QCL was tuned using the training data shown in panel (a). Outputs (*i.e.*, the predictive probability) for individual grid points are shown. (c) Predictive probability from QCLL after training. The formats and procedures are the same as those in panel (b).

Next, we treated a classification task. Following the numerical simulations in Mitarai et al. [26], the simple nonlinear binary classification task shown in Figure 3(a) was treated. The number of training samples was 200 (100 for class 1 and 100 for class 2). To treat this binary classification problem, two Hermitian matrices (*i.e.*, observables) $B_1$ and $B_2$ were used in step 3 of both algorithms. Here, a matrix whose first five diagonal entries are one and whose other entries are zero was used as $B_1$. A matrix whose 6–10-th diagonal entries are one and whose other entries are zero was used as $B_2$. The softmax function was used as the predefined function $F(\cdot)$. The cross-entropy function was used as the loss function $E(\cdot,\cdot)$. The number of qubits used to encode each input dimension was set to three, and the depth of the quantum circuit was set to three. The same optimization method as used in the previous regression task was used. The results are shown in Figure 3 (b) and 3(c). QCL and QCLL show similar behavior for this classification task.

Finally, using the same regression task, where $f(x) = x^2$ was estimated, we investigated the dependency of the models on the amount of training data and their robustness to noise. In the analysis to investigate the dependency of the models on the amount of training data, the number of training samples was changed within the range from 10 to 100. In the analysis to investigate their robustness to noise, the models were trained with training data including Gaussian noise. The training data were generated in the same manner as in the previous regression analysis except that the target outputs $y_n$ were set to $f(\mathbf{x}_n) + \varepsilon_n$ where $\varepsilon_n$ followed a Gaussian distribution with a mean of zero and a standard deviation of $\sigma$. $\sigma$ was varied within the range from 0.0 to 0.45. To quantitatively evaluate the prediction accuracy of a model trained with a given training data, we computed the root mean squared error (RMSE) between $f(\mathbf{x})$ and the model prediction across points in $[-1.0, 1.0]$. As done in the Appendix of Mitarai et al. [26], as another machine-learning algorithm to be compared, we trained a simple linear regression model (*i.e.*, ordinary least squares) using $\left\{x^6, x^5\sqrt{1-x^2}, x^4(1-x^2), x^3\sqrt{1-x^2}^3, x^2(1-x^2)^2, x\sqrt{1-x^2}^5, (1-x^2)^3\right\}$ as basis functions; these being polynomials appearing in $|\psi_{\text{in}}(\mathbf{x})\rangle$ of QCL and

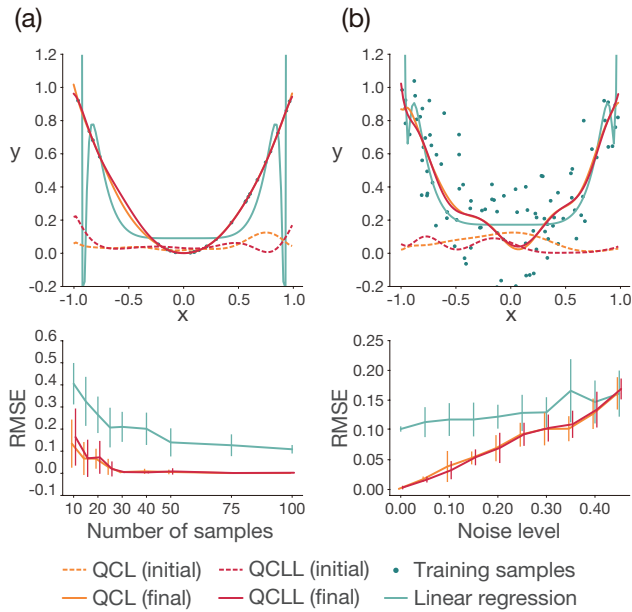

QCL (initial) ---- QCLL (initial) • Training samples
QCL (final) —— QCLL (final) —— Linear regression

**FIG. 4.** Dependency of the models on the amount of training data and their robustness to noise. (a) Dependency on the amount of training data. QCL, QCLL, and a simple linear regression model were trained with a small number of samples ($N = 25$, top). The formats are the same as in Figure 2. The root mean squared error (RMSE) averaged across 10 simulation repetitions is plotted as a function of the number of training samples (bottom). (b) Robustness to noise. The three regression models were trained with training data including Gaussian noise ($\sigma = 0.2$, top). The formats are the same as in Figure 2. The RMSE averaged across 10 simulation repetitions is plotted as a function of the noise level (bottom).

$\hat{\mathbf{v}}_{\text{in}}(\mathbf{x})$ of QCLL. As in Mitarai et al. [26], we examined how much QCL and QCLL avoided the risk of overfitting compared to this simple regression model.

To compare the dependency of the models on the amount of training data used, we evaluated the RMSE of each algorithm while changing the number of training samples (Figure 4(a)). In addition, to compare the robustness of the models to noise, we evaluated the RMSE of each algorithm while changing the level of Gaussian noise (Figure 4(b)). In both analyses, QCL and QCLL showed similar RMSEs, both of which were much lower than that of the simple linear regression. Those results suggest that QCL and QCLL efficiently avoid the risk of overfitting in similar manners.

### B. Behavior for real benchmark datasets

We evaluated the prediction performance of QCL, QCLL, and the linear regression using several benchmark datasets. We conducted almost the same experiment as shown in Figure 4(a) while replacing the simulation data with data from real benchmark datasets. Three representative benchmark datasets, Boston house-prices dataset [48], Diabetes dataset [49], and Iris flower dataset [50] were used. We performed regression analysis using the first two datasets, and performed classification analysis using the third dataset. To evaluate the prediction performance for independent test data, we divided each dataset into two groups: training data and test data. We randomly selected 20% of the samples in each dataset and used them as the test data, with the remaining data used as training data. The prediction error was quantified based on the root mean squared error (RMSE) for regression analysis and on the percentage of incorrect answers for classification analysis. To examine the ability to prevent overfitting, each machine-learning model was fitted (trained) with randomly selected $r$% of the training data, and the prediction error was then computed with the test data. The value of $r$ was changed across the range of {10, 20, 30, 40, 50, 75, 100}. Boston house-prices, Diabetes, and Iris flower datasets consist of 13, 10, and 4 input features (input variables), respectively. We picked up a pair of input features and those were fed to each machine-learning model. The prediction error was computed for each of all possible pairs in each dataset. For QCL and QCLL, the number of qubits used to encode each input feature was set to three and the depth of the quantum circuit was set to six. The other settings and procedures are the same as the previous regression or classification analysis. The linear regression was also trained using the basis functions (polynomials) that appear in QCL and QCLL. The RMSE and classification error were shown as functions of the proportion of the training data used (Figure 5). QCL and QCLL show better prediction performance compared with the simple linear regression.

### C. Complementary analysis to examine the approximation performance of QCLL

To quantitatively evaluate the ability of QCLL to approximate QCL, we performed a complementary analysis. In this analysis, we computed the proportion of the functions that QCLL can approximate among all the functions represented by QCL as follows:

1) Prepare a QCL model in which the learnable parameters (i.e., $\boldsymbol{\theta}$) and unitary transformations specifying the parameterization (i.e., $U_1, \cdots, U_M$ in Eq. (5)) are randomly set. Denote the function that this QCL model represents by $f_{QCL}(x)$.

2) Using the above QCL model, prepare a dataset $\mathcal{D} = \{(x_n, y_n)\}_{n=1}^{100}$ where $x_1, x_2, \cdots, x_{100}$ are distributed in [-1,+1] with equal spacing and $y_n = f_{QCL}(x_n)$.

3) Prepare a QCLL model in which the count sketch matrices are randomly set.

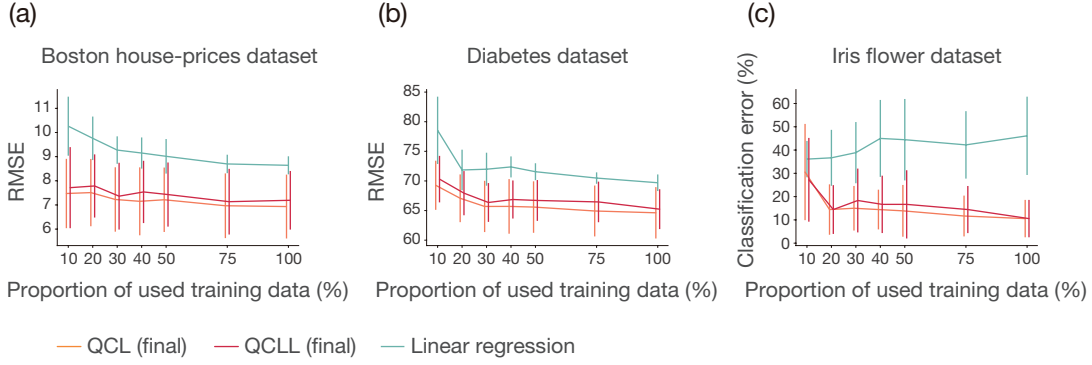

**FIG. 5.** Prediction performance for benchmark datasets. (a) Prediction performance for Boston house-prices dataset. QCL, QCLL, and a simple linear regression model were trained and tested using this dataset. The prediction error averaged across different input features was shown as a function of the amount of the training data. Error bars indicate the standard deviation across input features. (b) Prediction performance for Diabetes dataset. The formats are the same as in (a). (c) Prediction performance for Iris dataset. The formats are the same as (a) except that the classification error is plotted.

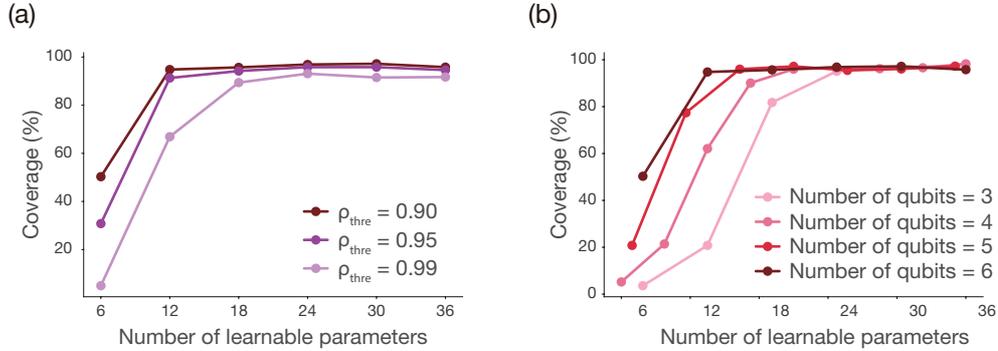

**FIG. 6.** Approximation performance of QCLL. (a) Approximation performance of QCLL as a function of the number of learnable parameters. To evaluate how well QCLL can approximate QCL, we compute the proportion of the functions that QCLL can approximate with an approximation accuracy higher than a predefined threshold $\rho_{thre}$ among all the functions represented by QCL. The resultant proportion value is referred to as "coverage." The coverage is plotted with three different predefined thresholds ($\rho_{thre}$ = 0.90, 0.95, and 0.99) as a function of the number of learnable parameters. The other settings are the same as in Figure 2 (The number of qubits in QCL is set to 6). (b) Approximation performance of QCLL for QCL with different numbers of qubits. The coverage is computed while the number of qubits in QCL is varied from 3 to 6. The coverages for the different numbers of qubits are shown in the same format as in (a). $\rho_{thre}$ is set to 0.90.

4) Fit the above QCLL model to $f_{QCL}(x)$ by optimizing the learnable parameters of the QCLL model using $\mathcal{D}$ as the training data. Denote the resultant function by $f_{QCLL}(x)$.

5) Compute the Pearson correlation coefficient between $f_{QCL}(x)$ and $f_{QCLL}(x)$ for $x$ distributed in [-1,+1]. Denote the resultant correlation value by $\rho$.

6) Repeat Steps 1–5 a sufficient number of times (1000 times) and compute the probability that $\rho$ is larger than a predefined threshold $\rho_{thre}$ (*e.g.*, 0.90).

We referred to the probability value computed in the above Step 6) as "coverage" in this study, and this value indicates the proportion of the functions that QCLL can approximate with an approximation accuracy higher than the predefined threshold among all the functions represented by QCL. For example, if the coverage is 100%, it indicates that QCLL can approximate any functions represented by QCL. For all numerical simulations reported here, the number of learnable parameters was matched between QCL and QCLL, and this number was varied. While setting the other settings to the same as in Figure 2, we plotted the coverage as a function of the number of learnable parameters (Figure 6(a)). The coverage monotonically increased with an increase in the number of learnable parameters, and the coverage reached more than 90%, even for a high predefined threshold ($\rho_{thre}$ = 0.99). We also examined whether this tendency was consistently observed when varying the number of qubits (*i.e.*, $Q$) in QCL. In the settings adopted in Figure 2, the number of qubits was set to six. Here, we varied it from three to six and plotted the coverages in the same manner (Figure 6(b)). The coverage monotonically increased when the number of learnable parameters increased in all cases. We also found that the coverage monotonically increased when the number of qubits increased. These results suggest that, although QCLL cannot approximate all the functions represented by QCL, QCLL can approximate a large proportion of them, especially when QCL consists of a large number of qubits and a large number of learnable parameters.

## IV. DISCUSSION AND CONCLUSIONS

We proposed a classical machine-learning algorithm that shares several properties with quantum circuit learning (QCL). Specifically, our proposed algorithm, quantum circuit-like learning (QCLL) internally uses the same $2^Q$-dimensional Hilbert space as that used in QCL. Despite this property, QCLL can be run on a classical computer with a low computational cost. Both algorithms also share the property of maintaining the norm of the weight vector (*i.e.*, coefficient vector) at one, which mitigates the risk of overfitting. The numerical simulations show that QCLL has similar behavior and performance to QCL on the machine-learning tasks treated in the original QCL study. This provides a new perspective from which to consider the advantages of QCL in terms of these properties.

While we demonstrated that QCLL shows similar prediction performance to QCL in several machine-learning tasks, it should be noted that these results are consistent with the expectations from previous quantum machine learning studies. Using the framework of probably approximately correct (PAC) learning, previous studies compared the number of training samples to achieve a given prediction performance between quantum and classical machine-learning algorithms with the same Vapnik–Chervonenkis (VC) dimension, and proved that their difference is only up to polynomial with respect to their VC dimension [41–44]. Thus, theoretically, for any given quantum machine-learning algorithm, there exists a classical machine-learning algorithm with similar prediction performance. In our numerical experiments, the prediction error of QCL was similar to, but slightly lower than that of QCLL (Figures 4 and 5). Because the above mentioned PAC learning framework assumes that the training data are noiseless, this observed difference might be explained by the effect of noise on the training data [51], as well as the polynomial factor expected by the PAC learning framework.

Similarly, our results do not contradict recent literature on time complexity because 1) QCL's computational advantage over classical algorithms has not been clear or proven and 2) even for quantum machine-learning algorithms that had been believed to have exponential speedup (*e.g.*, quantum algorithm for linear equations [52], quantum principal component analysis [53], and quantum recommendation systems [54]), classical counterparts only polynomially slower than those algorithms were recently proposed [55–59]. An attempt to construct a computationally efficient classical algorithm that directly approximates QCL may be helpful for further characterization of QCL.

In contrast to the quantum advantage in standard machine-learning tasks, a series of recent studies have suggested that quantum machine-learning algorithms can be exponentially more efficient than their classical counterparts in quantum state tomography [60] while acknowledging the power of the classical machine-learning algorithms [61,62]. The quantum advantage of QCL can also be characterized by assessing its efficiency in the context of quantum state tomography [63].

To make the random parameterization of the QCLL model more flexible, we can optionally use a random unitary matrix in QCLL by adopting Eq. (15). Such a modification might lead to better prediction performance in machine-learning tasks. We tested this in our preliminary experiments and found that the performance and behavior of QCLL with Eq. (15) are almost the same as those of the original QCLL using Eq. (14) (see APPENDIX). Thus, for simplicity, we adopted QCLL along with Eq. (14) in the main part of this study.

To approximate the inner product in the high-dimensional Hilbert space, we introduced two statistical techniques, count sketch and tensor sketch. For a similar purpose, several other approximation techniques have been developed in the machine learning field. The pioneering work by Rahimi and Recht [64] first presented a random projection-based algorithm to approximate the inner product associated with shift-invariant kernels (*e.g.*, Gaussian kernel). While their original method can be applied only to shift-invariant kernels, subsequent studies extended this technique to generalized radial basis kernels [65], polynomial kernels [66], and additive kernels [67,68]. Because random projections are also utilized in our method, our method can be regarded as an extension of the above line of work for the inner product of vectors in the form of Eq. (10).

While we introduced the count and tensor sketch techniques to create QCLL in this study, their application to quantum simulation might also be possible. For this purpose, although out of the scope of this study, investigating what class of quantum states can be efficiently treated by such sketch algorithms remains an important challenge, which may provide new insights into their properties.

## ACKNOWLEDGMENTS

The authors would like to thank Shinji Nishimoto and Chikako Koide for preparing the environment for the analysis, and Kazuho Watanabe and Kohei Hayashi for helpful comments on the manuscript.

## FUNDING

This research was supported by JSPS KAKENHI Grant number 20K16465, MEXT Quantum Leap Flagship Program (MEXT Q-LEAP) Grant Number JPMXS01 20330644, JST ERATO Grant Number JPMJER1801, and JST PRESTO Grant Number JPMJPR2128.

## COMPETING INTERESTS

All authors declare no competing financial interests.

## APPENDIX: EFFECT OF DIFFERENT CHOICES OF RANDOM PARAMETRIZATION

Using Eq. (15) instead of Eq. (14) in QCLL, we performed the analyses same as in Figures 2, 3, 4 and 6. The results are shown in Figures 7, 8, 9, and 10. Almost the same behavior and performance were observed for QCLL with Eq. (14) and QCLL with Eq. (15).

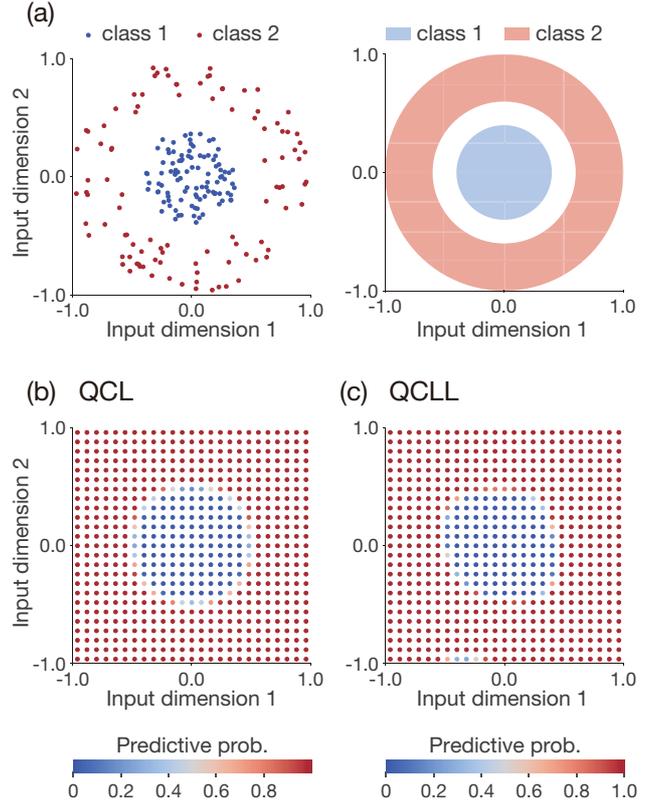

**FIG. 8.** Performance of the modified QCLL applied on a classification task. The formats and procedures are the same as in Figure 3.

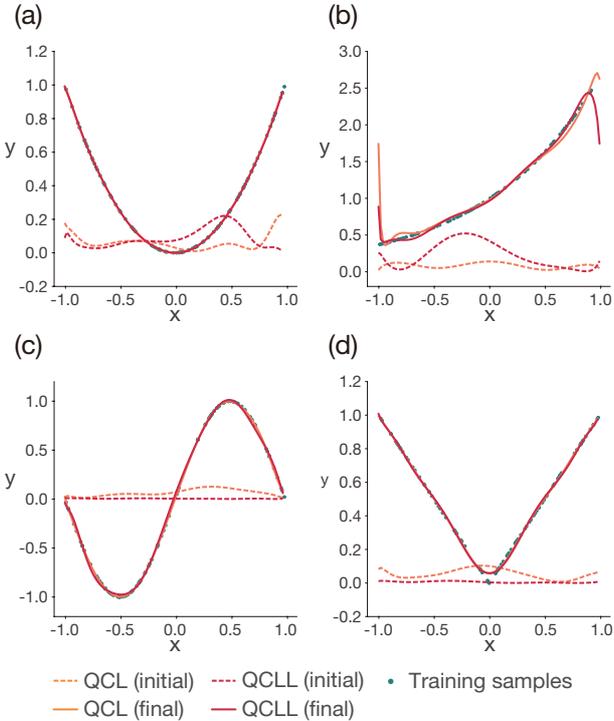

**FIG. 7.** Performance of modified QCLL on regression tasks. The formats and procedures are the same as in Figure 2.

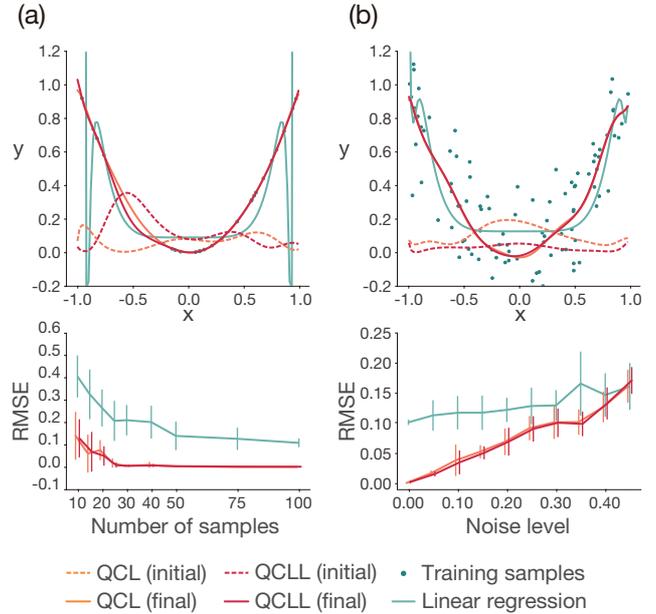

**FIG. 9.** Dependency of the modified QCLL on the amount of training data and its robustness to noise. The formats and procedures are the same as in Figure 4.

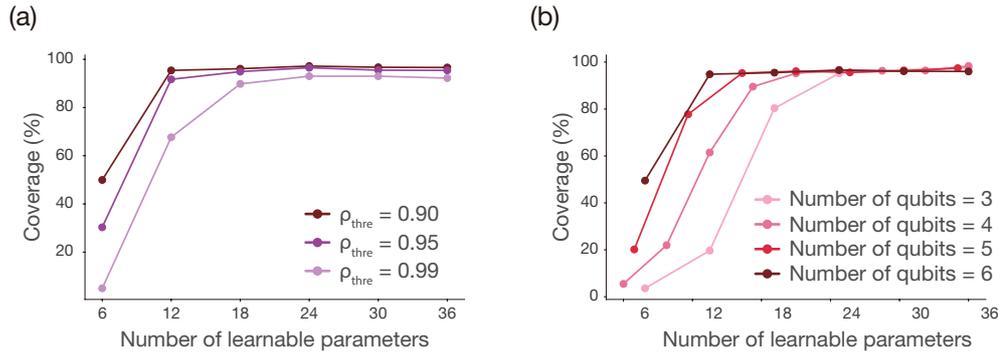

**FIG. 10.** Approximation performance of the modified QCLL. The formats and procedures are the same as in Figure 6.

bibliography**34**, 480 (2012).